\documentclass[12pt]{article}
\usepackage{a4} \textheight=22cm
\newcommand{\be}{\begin{equation}}\newcommand{\ee}{\end{equation}}
\newcommand{\beq}{\begin{equation}}\newcommand{\eeq}{\end{equation}}
\newcommand{\bea}{\begin{eqnarray}}\newcommand{\eea}{\end{eqnarray}}
\newcommand{\beao}{\begin{eqnarray*}}\newcommand{\eeao}{\end{eqnarray*}}

\newcommand{\ep}{{\epsilon}}\newcommand{\om}{{\omega}}
\newcommand{\plabel}[1]{\label{#1}}
\begin{document}
\begin{center}
{\bf \Large Path  integral quantization of electrodynamics}\\[4pt] 
{\bf \Large  in dielectric media}
\end{center}
\vspace{12pt}
\begin{center}
M. Bordag\footnote{http://www.physik.uni-leipzig.de/$\tilde{\quad}$bordag},
  K. Kirsten\footnote{e-mail: kirsten@itp.uni-leipzig.de}\\[5pt]
{\small Institute for Theoretical Physics, Leipzig
University\\Augustusplatz 10/11, 04109 Leipzig, Germany}\\[10pt]
 and \\[10pt]D.V. Vassilevich\footnote{e-mail: vassil@snoopy.niif.spb.su}\\
{\small Department of Theoretical Physics,
     St.Petersburg University\\
     198904 St.Petersburg, 
     Russia}
\end{center}
\abstract{In the present paper we study the Faddeev--Popov path integral
quantization of electrodynamics in an inhomogenious dielectric
medium. We quantize all polarizations of the photons and introduce the
corresponding ghost fields. Using the heat kernel technique, we
express the heat kernel coefficients in termini of the dielectricity
$\epsilon (x)$ and calculate the ultra violet divergent terms in the
effective action. No cancellation between ghosts and "non-physical"
degrees of freedom of the photon is observed.}

\section{Introduction}
The Casimir effect describes the
forces resulting from the vacuum fluctuations (ground state energy) of the
electromagnetic field in simple situations realized by conducting
surfaces. These forces can be viewed as retarded Van der Waals
forces between the atoms constituting the surfaces (and the bodies behind).
As a generalisation of this picture one can consider some medium. It can be
characterised either by atoms at positions $x_{i}$ with their individual
polarisabilities $\alpha_{i}$ or by a macroscopic permittivity $\ep(x)$ and
permeability $\mu(x)$. Again, we can calculate the resulting potential of 
the VanderWaals forces or the vacuum energy $E_{0}[\ep(x),\mu(x)]$ of the
electromagnetic field in a background given by $\ep(x)$ resp. $\mu(x)$.
Taking into account that real permittivity resp. permeability are functions of
the photon frequency we arrive at the problem to calculate
$E_{0}[\ep(x,\om),\mu(x,\om)]$. The dependence on $\om$ has as a physical
background, besides others the observation that any medium becomes transparent
for $\om$ sufficiently high (we do not consider inelastic effects here).
Therfore $\ep,\mu\to 1$ for $\om\to\infty$ should serve as a natural
ultraviolet regularisation. This is widely believed, but not shown in a
rigorous way yet.

The problem of the calculation of  
$E_{0}[\ep(x),\mu(x)]$, i.e. without frequency
dependence, may be well posed independently. A physical justification could be
that the essential contribution results after a proper renormalisation from
quite low frequencies $\om$, where $\ep$ and $\mu$ can be viewed as
approximately independent on $\om$. In that case we don't have a natural
regularisation and have to proceed like in the general situation with sharp
boundary conditions or a general background field. For technical reasons we
use the zeta-functional regularisation. Then the first step is to calculate
the divergent contributions (the proper technical tool being the heat kernel
expansion), the second is to formulate a model for the interpretation of the
renormalisation (this is to be able to reinterpret the subtraction of the
divergencies as a renormalization of classical quantities like volume, surface
tension etc. as discussed in \cite{wipf} or like mass and coupling constant of
the background field as discussed in \cite{bordagkirsten95}) and, finally, in
a third step to calculate the renormalised groundstate energy $E_{0}$.  In the
present paper we carry out the first step and discuss the second to some
extend.

The  forces resulting from the electromagnetic 
vacuum fluctuations in polarisable
media have been given much attention to. The common features of these
investigations are sharp boundaries separating regions of different values of
$\ep (x)$ and simple geometries (planes, cylinders
and so on). For instance, there
have been interesting calculations (mainly with respect to the sign of the
force) for a dielectric sphere \cite{brevik}. 
Also, much attention had been spent on a possible explanation of
sonoluminescence as a dynamical Casimir effect, especially in a series by
Schwinger \cite{schwinger}. Recently, the bulk and surface energy
contributions had been discussed \cite{visser,MilNg}.

However, with respect to the
renormalisation it is  difficult to deal with sharp
boundaries resp. non smooth background fields. It is known that additional
contributions to the heat kernel expansion occur and therefor  additional
counter terms result for which a general theory is still missing. Therefor we
restrict ourselves in the present paper to $\ep(x)$ which are smooth
functions on $x$.

There is still another problem we have to pay attention. In the common
understanding the quantisation of QED in media is done in the Coulomb
gauge, i.e., the two 'physical' polarisations of the photon are
quantised. Also there are known procedures where all polarisations of
the photon are quantised and the gauge invariance (in the presence of
boundaries) is restored by ghosts which have to fulfil boundary
conditions too (one of the first is \cite{ambjornhughes83}, later on
it had been discussed in \cite{esposito}). In most cases their
contributions cancel that resulting from the 'unphysical' photons, but
counter examples are known (e.g. for QED in curved space time,
\cite{DV}).  In the framework of quantum optics the canonical
quantization of photons was considered in \cite{Glauber} without,
however, analysing the ghost contributions. An alternative approach
for quantization in covariant gauge without ghosts, however restricted
to sharp boundaries, had been developed in \cite{borowi85}.

In the present paper we analyse the problem of QED with a position
dependent permittivity $\ep(x)$ from the point of view of general
quantum gauge theory in an external field.  We analyse the canonical
path integral measure and corresponding configuration space measure. A
gauge fixing term is introduced together with the ghost action.  Next
we analyse the ultra violet structure of the theory by means of the
heat kernel expansion. No cancellation between ghosts and photon modes
is obtained.

Our paper is organized as follows. In the next section the quantization of the
theory is considered and the path integral is derived. In sec. 3 we use the
heat kernel expansion to evaluate ultra violet divergencies.  Concluding
remarks are given in sec 4. 
An Appendix contains an alternative calculation to
check up the results of sec. 3.
 \section{Canonical quantization and gauge choice} Consider the action
for the electromagnetic field in a dielectric media with
 permittivity $\epsilon (x)$: \beq S=\int d^4x \frac 12 (\epsilon (x) E^2 -
 B^2 ) \plabel{act} \eeq To avoid technical complexities we put the
permeability
 $\mu =1$ and suppose that $\epsilon $ depends on spatial coordinates only.

Let us rewrite the action (\ref{act}) in the canonical first order
form: \beq S_1= \int d^4x (P^i \partial_0 A_i + A_0 \partial_i P^i -
\frac 1{\epsilon (x)} P^i P^i - \frac 12 B^2 ) \plabel{first} \eeq
Here $A_\mu$ is the vector potential. $P^i=-\epsilon (x) E_i$ is the
momentum conjugate to $A_i$.  Canonical Poisson brackets are
\begin{equation} \{ A_i({\bf x},t),P^j({\bf y},t)\}=\delta^j_i \delta
({\bf x}-{\bf y}) \end{equation} The same brackets were obtained in
\cite{Glauber}. $A_0$ plays the role of a Lagrange multiplier
generating the Gauss law constraint, which in turn generates gauge
transformations. According to the general method \cite{FS} of
quantization of gauge theories we can write down the path integral
\beq Z=\int DA_i DA_0 DP^j J_{FP}\delta (\chi (A_i))\exp (iS_1) ,
\plabel{Z1} \eeq where $\chi (A_i)$ is a gauge fixing condition,
$J_{FP}$ is the Faddeev--Popov determinant, $J_{FP}=\det \{ \chi (A),
\partial_jP^j\}$.  Now we can perform the integration over the momenta
$P^j$. It produces the factor $\prod_x \sqrt {\epsilon (x)^3}$ which
should be absorbed in the path integral measure $DA_i$. We arrive at
the following expression: \beq Z=\int D\tilde A_i DA_0 J_{FP} \delta
(\chi (A)) \exp (iS), \quad \tilde A_i =\sqrt \epsilon A_i .
\plabel{Z2} \eeq Our $\tilde A$ variables coincide with the $q'$
variables of Glauber and Lewenstein \cite{Glauber}.  Note that the
measure in (\ref{Z2}) differs from the naive one $\prod DA_\mu$.  We
can use the Faddeev--Popov trick to transform the path integral
(\ref{Z2}) to whatever gauge condition we prefer, introduce a gauge
fixing term and ghost fields.  There is nothing specific in this
respect in the present model.  All steps repeat those of a standard
text book \cite{FS}. The result is \bea Z&=&\int D\tilde A_i DA_0\ Dc\
D\bar c \exp \left\{i\int d^4x [ \frac 14 (2\epsilon (x) (\partial_0
\epsilon^{-1/2} \tilde A_i- \partial_i A_0)^2 \right.\nonumber \\ &\
&\left.- (\partial_i \epsilon^{-1/2} \tilde A_k - \partial_k
\epsilon^{-1/2} \tilde A_i )^2) +{\cal L}_{gf}+ {\cal L}_{ghost}
]\right\} \plabel{Z3} \eea where ${\cal L}_{gf}$ and ${\cal
L}_{ghost}$ are gauge fixing term and ghost action respectively.  As
usual, we can bring the action in (\ref{Z3}) to the form $\int A_\mu
L_{\mu\nu}A_\nu$, where $L_{\mu\nu}$ is a second order differential
operator. In calculating the effective action and the heat kernel
expansion it is much more convenient to deal with operators of Laplace
type, i.e. operators with scalar leading symbol. There is a unique
gauge choice which splits the $L_{\mu\nu}$ in a direct sum of
operators of Laplace type. This choice is \begin{equation} {\cal
L}_{gf}=-\frac 12 (\epsilon^{-1} \partial_i \epsilon^{1/2} \tilde A_i
- \epsilon \partial_0 A_0 )^2 \plabel{Lgf} \end{equation}
\begin{equation} {\cal L}_{ghost}= -\bar c (-\epsilon^{-1} \partial_i
\epsilon \partial_i + \epsilon \partial_0^2 )c \nonumber
\end{equation} The action for the electromagnetic field $A$ then takes
the form \bea & &\frac 12 \int d^4x[\epsilon (\partial_i A_0)^2
-\epsilon^2(\partial_0A_0)^2 +(\partial_0\tilde A_i)^2
 \\
& &+\tilde A_i\epsilon^{-1/2} 
(\partial_j^2\delta_{ik}
-e_i\partial_k+\partial_ie_k-e_
ie_k)\epsilon^{-1/2}
\tilde A_k], \qquad 
e_i=\partial_i {\rm ln}\epsilon
\nonumber\eea
Note, that the 
mixing between $A_0$ and 
$\tilde A_i$ is removed 
completely.

The total action with gauge fixing 
and ghost term is invariant 
under
the BRST transformations with the  
parameter $\sigma (x)$:
\bea
\delta A_0 &= &\partial_0 
\sigma c \nonumber \\
\delta \tilde A_i &= 
&\epsilon^{1/2} \partial_i 
\sigma c \nonumber \\
\delta c &= &0 \label{BRST} \\
\delta \bar c &= & 
(-\epsilon^{-1} \partial_i 
\epsilon^{1/2} \tilde A_i+
\epsilon \partial_0 A_0 ) 
\sigma \nonumber
\eea
which are given here to 
complete the picture.

\section{Effective action and 
heat kernel expansion}

Now we are able to integrate over $A_0$, $\tilde A$ and the
ghosts. The resulting path integral reads after Wick rotation to the
Euclidean domain: \beq Z=Z[A_0]Z[\tilde A]Z[\bar c,c] ,\plabel{Z4}
\eeq where the separate contributions are of the form: \bea Z[A_0]&=&
{\det }^{-1/2} (-\partial_i \epsilon \partial_i -\epsilon^2
\partial_0^2 ) \nonumber \\ Z[\tilde A]&=& {\det }^{-1/2} \left (
-\frac 1\epsilon \partial_k^2\delta_{ij} -\partial_0^2\delta_{ij}
-G_i\partial_j+G_j\partial_i-M_ {ij} \right ) \plabel{dets} \\ Z[\bar
c,c]&=& \det (-\epsilon^{-1}\partial_i \epsilon \partial_i - \epsilon
\partial_0^2 ) \nonumber \eea We introduced the notations:
\begin{equation} G_i=\frac {e_i}\epsilon \qquad M_{ij}=\frac 1\epsilon
(e_{ij}-e_ie_j) \qquad e_{ij}=\partial_i e_j \end{equation} For the
functional determinants we use the integral representation \beq \log
\det (L)= \int_0^\infty \frac {dt}t K(L;t) \plabel{irep} \eeq where
the heat kernel $K(L;t)$ for a second order elliptic operator $L$ is
\beq K(L;t)={\rm Tr} \exp (-tL) \plabel{hk} \eeq

The ultraviolet behavior of functional determinants is given by the
asymptotic expansion of the heat kernel (\ref{hk}) as $t\to +0$. Since
all the operators are of Laplace type, we can use the general theory
\cite{Gil}. Each of the operators has the structure \begin{equation}
L=-(g^{\mu\nu}\partial_\mu\partial_\nu +a^\sigma \partial_\sigma +b)
\plabel{gen} \end{equation} where $g^{\mu\nu}$ plays the role of a
metric. $a^\sigma$ and $b$ are local sections of endomorphism ${\rm
End}(V)$ of certain vector bundle. By introducing a connection
$\omega_\mu$ in the vector bundle $V$, one can bring $L$ to the form:
\begin{equation} L=-(g^{\mu\nu}\nabla_\mu\nabla_ \nu +E) \plabel{gen1}
\end{equation} where $\nabla$ is a sum of the Riemannian covariant
derivative with respect to the metric $g$ and the connection $\omega$.
The explicit form of $\omega$ and $E$ is \begin{eqnarray}
\omega_\delta&=&\frac 12 g_{\nu\delta}(a^\nu
+g^{\mu\sigma}\Gamma_{\mu\sigma }^\nu ) \nonumber \\ E&=&b-g^{\mu\nu}
(\partial_\mu\omega_\nu + \omega_\mu \omega_\nu - \omega_\sigma
\Gamma^\sigma_{\mu\nu} ) \plabel{omega} \end{eqnarray} As usual,
$\Gamma$ denotes the Christoffel connection.

Given the geometric quantities $g$, $\omega$ and $E$, we are able to
calculate the coefficients $a_n$ of the asymptotic expansion
\begin{equation} {\rm Tr} (f\exp (-tL))=t^{-2}\sum_{n=0}^\infty t^n
a_n (f,L) \plabel{hf} \end{equation} for a function $f$. The
coefficients $a_n (f,L)$ contain information on the asymptotics of the
heat kernel diagonal $<x\vert \exp (-tL) \vert x>$.  The analytical
expressions for the first coefficients are known \cite{Gil}:
\begin{eqnarray} a_0&=& \frac 1{(4\pi)^2} {\rm tr}_V \int d^4x g^{1/2}
f \nonumber \\ a_1&=& \frac 1{(4\pi)^2} {\rm tr}_V \int d^4x g^{1/2}
f(E+ \frac \tau{6}) \nonumber \\ a_2&=& \frac 1{(4\pi)^2} {\rm tr}_V
\int d^4x g^{1/2} \frac 1{360} f (60{E_{;\mu}}^\mu+60\tau E +180
E^2+30\Omega_{\mu\nu} \Omega^{\mu\nu} \nonumber \\ &\ &
+12{\tau_{;\mu}}^\mu +5\tau^2-2\rho^2+2R^2 ) \plabel{an}
\end{eqnarray} Here $R$, $\rho$ and $\tau$ are Riemann tensor, Ricci
tensor and scalar curvature of the metric $g$ respectively.  Semicolon
denotes covariant differentiation, $E_{;\mu}=\nabla_\mu E$.  All
indices are lowered and raised with the metric tensor, ${\rm tr}_V$ is
the bundle (matrix) trace, $\Omega$ is the field strength of the
connection $\omega$: \begin{equation} \Omega_{\mu\nu}=\partial_\mu
\omega_\nu -\partial_\nu \omega_\mu +\omega_\mu \omega_\nu -
\omega_\nu \omega_\mu \end{equation}

The three coefficients (\ref{an}) are enough to describe the one--loop
ultra violet divergencies in four dimensional quantum field theory in
an infinite space--time.

Now our problem is reduced to the calculation of the geometric
quantities appearing in (\ref{an}).  For the ghost operator we have
\bea g_{ij}&=&\delta_{ij} ,\quad g_{00}=\epsilon^{-1}(x) \nonumber \\
\Gamma_{00}^i&=&\frac 1{2\epsilon}e_i \quad \Gamma_{0i}^0=-\frac 12
e_i \nonumber \\ \omega_0&=&0,\quad \omega_i=\frac 34 e_i \nonumber \\
E&=&-\frac 34 e_{ii}-\frac 3{16} e_ie_i \nonumber \\ {R^i}_{jkl}&=&0
\plabel{geogh} \\ {R^0}_{i0j}&=&-\frac 12 e_{ij}+\frac 14 e_ie_j
\nonumber \\ \rho_{ij}&=&\frac 12 e_{ij}-\frac 14 e_ie_j \nonumber \\
\rho_{00}&=&\frac 1{2\epsilon} (e_{ii}-\frac 12 e_je_j) \nonumber \\
\tau &=& e_{ii}-\frac 12 e_ie_i \nonumber \eea For the operator acting
on $A_0$ the relevant quantities are: \bea
g_{ij}&=&\epsilon^{-1}\delta_{i j} ,\quad g_{00}=\epsilon^{-2}(x)
\nonumber \\ \Gamma_{00}^i&=&\frac 1{\epsilon} e_i \quad
\Gamma_{0i}^0=- e_i \nonumber \\ \Gamma_{ij}^k&=&-\frac 12
(e_i\delta_{jk}+e_j\delta_{ik}- e_k\delta_{ij} ) \nonumber \\
\omega_0&=&0,\quad \omega_i=\frac 54 e_i \nonumber \\ E&=&\epsilon
\left (-\frac 54 e_{ii}+\frac 5{16} e_ie_i \right ) \nonumber \\
{R^i}_{jkl}&=&\frac 12 (-e_{jl}\delta_{ik}+e_{jk}\delta_{il}
+e_{il}\delta_{kj}-e_{ik}\delta _{lj}) \plabel{geoA0} \\ &\ &+\frac 14
(e_pe_p(\delta_{jl}\delta_{ki}- \delta_{jk}\delta_{li} )
+e_ke_j\delta_{li}-e_ke_i\delta _{jl}-e_le_j\delta_{ki}
+e_le_i\delta_{jk}) \nonumber \\ {R^0}_{i0j}&=&- e_{ij}+\frac 12
e_le_l \delta_{ij} \nonumber \\ \rho_{ij}&=&\frac 32 e_{ij}+\frac 12
\delta_{ij}e_{kk} +\frac 14 e_ie_j-\frac 34 \delta_{ij}e_ke_k
\nonumber \\ \rho_{00}&=&\frac 1{\epsilon} (e_{ii}-\frac 32 e_je_j)
\nonumber \\ \tau &=&\epsilon (4e_{ii}-\frac 72 e_ie_i )\nonumber \eea
For the operator acting on $\tilde A$ we obtain: \bea
g_{ij}&=&\epsilon\delta_{ij} ,\quad g_{00}=1 \nonumber \\
\Gamma_{ij}^k&=&\frac 12 (e_i\delta_{jk}+e_j\delta_{ik}-
e_k\delta_{ij} ) \nonumber \\ \omega_l^{ab}&=&\frac 12
(-e_a\delta_{bl}+e_b\delta_{al} -\frac 12 e_l\delta_{ab} ) \nonumber
\\ E_{ab}&=&M_{ab}+\frac 1{4\epsilon} (e_{kk}\delta_{ab}+e_ae_b+\frac
54 e_pe_p\delta_{ab} ) \plabel{geoA} \\ {R^i}_{jkl}&=&\frac 12
(e_{jl}\delta_{ik}-e_{jk}\delta _{il} -e_{il}\delta_{kj}+e_{ik}\delta
_{lj}) \nonumber \\ &\ &+\frac 14 (e_pe_p(\delta_{jl}\delta_{ki}-
\delta_{jk}\delta_{li} ) +e_ke_j\delta_{li}-e_ke_i\delta
_{jl}-e_le_j\delta_{ki} +e_le_i\delta_{jk}) \nonumber \\
\rho_{jk}&=&-\frac 12 (e_{jk}+e_{pp}\delta_{jk}) +\frac 14
(e_ke_j-e_pe_p\delta_{kj} ) \nonumber\\ \tau &=&\frac 1{\epsilon}
(-2e_{pp}-\frac 12 e_pe_p) \nonumber \eea Here for convenience we
prefere to keep the distinction between coordinate indices $\{ i,j,k,l
\}$ and bundle indices $\{ a,b\}$, though they all run from 1 to 3.
In the equations (\ref{geogh}) - (\ref{geoA}) repeated indices are
contracted with the flat space metric $\delta_{ij}$.

It is instructive to express the heat kernel coefficients in terms of
$\epsilon$ and its derivatives: \bea K_{gh}(f,t) &=& \frac 1{(4\pi
t)^2} \int d^4x \epsilon^{-1/2}f \{ 1+t \left ( -\frac {7}{12} e_{ii}
-\frac {13}{48} e_ie_i \right ) \nonumber \\ \ &\ & +\frac {t^2}{360}
( -33 e_{iijj} -18 e_ie_{ijj} -33 e_{ij}e_{ij} +\frac {237}{4}
e_{ii}e_{jj} \nonumber \\ \ &\ & +\frac {531}{8} e_{ij}e_ie_j +\frac
{33}{4} e_{ii}e_je_j +\frac {837}{64} e_ie_ie_je_j ) +O(t^3) \}
\nonumber \\ K_{[A_0]}(f,t) &=& \frac 1{(4\pi t)^2} \int d^4x
\epsilon^{-5/2}f \{ 1+t \epsilon \left ( -\frac {7}{12} e_{ii} -\frac
{13}{48} e_ie_i \right ) \nonumber \\ \ &\ & +\frac
{t^2\epsilon^2}{360} ( -27 e_{iijj} -60 e_ie_{ijj} -41 e_{ij}e_{ij}
+\frac {119}{4} e_{ii}e_{jj} \nonumber \\ \ &\ & -\frac {91}{8}
e_{ij}e_ie_j +\frac {415}{8} e_{ii}e_je_j +\frac {4141}{64}
e_ie_ie_je_j ) +O(t^3) \} \nonumber \\ K_{[\tilde A]}(f,t)&=& \frac
1{(4\pi t)^2} \int d^4x \epsilon^{3/2}f \{ 1+\frac t\epsilon \left (
\frac 34 e_{ii} -\frac 1{16} e_ie_i \right ) \plabel{asym}\\ &\ &
+\frac {t^2}{360\epsilon^2} ( 81 e_{iijj} -111 e_ie_{ijj} +162
e_{ij}e_{ij} -\frac {711}{4} e_{ii}e_{jj} \nonumber \\ \ &\ & -\frac
{1029}{4} e_{ij}e_ie_j +\frac {793}{8} e_{ii}e_je_j +\frac {4263}{16}
e_ie_ie_je_j ) +O(t^3) \} \nonumber \eea Here $e_{i\dots j}=\partial_i
\dots \partial_j {\rm ln} \epsilon$. This completes the calculation of
the UV divergent terms.

 We can define a "total" heat kernel as $K_{[A_0]}+ K_{[\tilde
A]}-2K_{gh}$. We see, that the contribution of ghosts is not cancelled
by that of $A_0$ and of the "non--physical" components of $\tilde A$.

As a check, in the Appendix we derive (\ref{asym}) by an alternative
method.

The asymptotic expansion constructed above gives $2n$ spatial
derivatives of $\epsilon$ in any $a_n$. Hence it is clear that certain
smoothness of $\epsilon (x)$ is needed.  Our expansion is not valid if
$\epsilon$ changes abruptly, as e.g.  for a bubble in water. For the
configurations of latter type boundary terms in the heat kernel
expansion should be taken into account.

\section{Conclusions and  discussion} 
In the present paper we performed the path integral
quantization of electromagnetic fields in a dielectric medium. As a
first step, we considered the first order action and derived the
canonical Poisson brackets. Next, we constructed the canonical
(simplectic) measure in the phase space. We built up a measure in the
configuration space by means of an integration over the canonical
momenta. This measure appeared to be different from the naive one. By
choosing a suitable gauge fixing condition (\ref{Lgf}) we reduced the
path integral to a product of three determinants of operators of
Laplace type. For the evaluation of the ultra violet divergent parts
of this determinants the standard heat kernel technique \cite{Gil} is
available.  Our results are re-checked by another technique (see
Appendix).  We observed no cancellation of ultra violet divergencies
between ghosts and any "non-physical" components of the vector
potential.  Thus it is highly unlikely that the full quantized
electrodynamics in dielectric media is equivalent to a theory where
only two polarizations of photons are quantized.

The next step to do is to work out a suitable cut--off procedure for
the path integral. This problem is very non--trivial in the present
case. Since $\epsilon \to 1$ at high frequencies, the cut-off is {\it
physical}, it will not be removed after a renormalization.  Therefore,
we must be sure that the basic properties of the quantum field theory,
as unitarity and absence of gauge anomaly, are valid at finite
cut--off. After having solved this problem, it will be possible to
consider the vacuuum energy densities and other physical quantities of
interest.

\section*{Acknowledgments} Work of one of the authors (D.V.) was
partially supported by the Russian Foundation for Fundamental
Research, grant 97-01-01186, and by funds provided by Saxonia
Government. KK has been supported by the DFG under contract number Bo
1112/4-2.  \section*{Appendix} In this Appendix we describe briefly an
alternative method for the evaluation of the heat kernel expansion
which we used to control our results.

We can represent the functional trace in the r.h.s. of (\ref{hk}) as
an integral over $x$ of diagonal matrix elements between $<x|$ and
$|x>$ and insert "unity" expressed via an integral of momentum
eigenstates: \beq {\rm Tr}\exp (-tL) =\int \frac {d^4xd^4k}{(2\pi )^4}
<x|\exp (-tL) |k><k|x> \plabel{tr} \eeq The generic form of the matrix
element in (\ref{tr}) is $<x|F_1(\epsilon , \partial \epsilon ) F_2
(\partial )|k>$, where $F_1$ and $F_2$ are some polynomials of
$\epsilon$ and its derivatives and of $\partial_i$
respectively. Acting on the left $F_1$ is replaced by its value in the
point $x$. Acting on the right, $F_2$ is replaced by $F_2 (ik)$. It is
easy to see that the result is \beq \int \frac {d^4xd^4k}{(2\pi )^4}
\exp (-t L(\epsilon (x), \partial_\mu \to \partial_\mu +ik_\mu ))
\plabel{mat} \eeq where we should take all external fields in the
point $x$, shift all derivatives by $ik$, and drive derivatives to the
right.  It is understood, that $\partial$ standing at the very right
position vanishes.

Consider the heat kernel for the ghost operator: \bea K_{gh}(t)&=&\int
\frac {d^4xd^4k}{(2\pi )^4} \exp (t(\partial_i^2+ 2ik_j\partial_j +
\nonumber \\ & & +(\partial_j \log \epsilon )\partial_j + i k_j
(\partial_j \log \epsilon ) -k^2-\epsilon \omega^2 )) \plabel{hk-gh}
\eea where $\{k_\mu \} = \{\omega ,k_j\}$. Time derivatives are
dropped out because $\epsilon (x)$ is static.

To obtain a small $t$ asymptotic expansion of (\ref{hk-gh}), one
should isolate the factor $\exp (-t(k^2+\epsilon \omega^2))$ and
expand the rest of the expression in a power series of operators and
functions involved. Next one should integrate over momenta and collect
all terms with the same powers of proper time $t$. Denote the
exponential in (\ref{hk-gh}) as $\exp (A+B)$ , where $A=-t(k^2
+\epsilon \omega^2 )$. Note, that $A$ does not commute with $B$.
However, the repeated commutator $[[[B,A],A],A]$ vanishes. This allows
us to present the exponential as follows (see e.g. \cite{Yaj}) \bea
\exp (A+B) &=& \exp A (1+B+\frac 12 [B,A] + \frac 16 [[B,A],A]+
\plabel{AB} \\ &+&\frac 12 B^2 +\frac 12 [B,A]B +\frac 16 [B,[B,A]] +
\frac 18 [B,A]^2+\dots \nonumber \eea We retained all the terms which
contribute to the two leading terms of the asymptotic expansion
proportional to $t^{-2}$ and $t^{-1}$.

Acting as explained above we obtain the asymptotic expansions for the
heat kernels $K_{gh}$ , $K_{[A_0]}$ and $K_{[\tilde A]}$.  The first
two terms are in complete agreement with (\ref{asym}).  Calculations
of the third terms are too complicated to be done just for a control.


\begin{thebibliography}{99} \bibitem{wipf} S.K. Blau, M. Visser and
A. Wipf, Nucl. Phys. B {\bf 310} (1988) 163.
\bibitem{bordagkirsten95}M. Bordag and K. Kirsten, Phys. Rev. D {\bf
53} (1996) 5753.  \bibitem{brevik}I. Brevik, I. Skurdal and R. Sollie,
J. Phys. A, {\bf 27} (1994) 6853.  \bibitem{schwinger}J. Schwinger,
Proc. Nat. Acad. Sci, {\bf 90} (1993) 2105, 4505, 7285.
\bibitem{visser}C. Carlson, C. Molina-Paris, J. Perez-Mercader and
M. Visser, Phys. Lett. {\bf B 395} (1997) and hep-th/9707073.
\bibitem{MilNg}
 K.A.Milton and Y.J. Ng, Phys. Rev. E {\bf 55} (1997) 4207.
\bibitem{ambjornhughes83}J. Ambjorn and R.J. Hughes, Nucl. Phys. {\bf
B217} (1983) 336.  \bibitem{esposito}G. Esposito, A.  Kamenshchik and
G. Pollifrone, {\it Euclidean quantum gravity on manifold with
boundary}, Kluwer, 1997.\\ G. Esposito, A. Yu. Kamenshchik,
I. V. Mishakov, G. Pollifrone, Class. Quantum Grav. {\bf 11} (1994)
2939.\\ G. Esposito, A.Yu. Kamenshchik, I.V.  Mishakov and
G. Pollifrone, Phys. Rev. {\bf D 52} (1995) 2183.
\bibitem{DV}D.V. Vassilevich {Phys. Rev.} {\bf D 52} (1995) 999.
\bibitem{borowi85}M. Bordag, D. Robaschik, E. Wieczorek,
Ann. Phys. (N.Y.) {\bf 165} (1985) 192.  \bibitem{Glauber}R.J. Glauber
and M. Lewenstein, { Phys. Rev.} {\bf A 43} (1991) 467.
\bibitem{FS}L.D. Faddeev and A.A. Slavnov, {\it Gauge Fields:
Introduction to Quantum Theory}, Benjamin/Cummings, New York,1980.
\bibitem{Gil}P.B. Gilkey, {\it J. Diff. Geom.} {\bf 10} (1980) 601;\\
P.B. Gilkey, {\it Invariance theory, the heat equation and the
Atyah--Singer index theorem} (2nd edition), CRC Press, Boca Raton,
Florida, 1994.  \bibitem{Yaj}S. Yajima, {\it Class. Quantum Grav.}
{\bf 13} (1996) 2423.  \end{thebibliography}
 \end{document}